\documentstyle[aps,epsf]{revtex}
\begin{document}
\title{Magnetic Properties of High Temperature Superconductors: a Spin Liquid
Approach. \\ Talk given at Stanford Spectroscopy Meeting, 1995.}
\author{L. B. Ioffe}
\address{Physics Department, Rutgers University, Piscataway, NJ 08855}
\author{A. J. Millis}
\address{AT\&T Bell Laboratories, 600 Mountain Ave, Murray Hill NJ
07974}
\maketitle
\begin{abstract}
I argue that strongly correlated two dimensional electrons form a spin
liquid in some regimes of density and temperature and give the theory of the
magnetic properties of this spin liquid using the representation in terms of
femions interacting with a gauge field.
I show that this state is characterized by anomalous power law spin
correlations and discuss the implications of these correlations for the
temperature dependence of NMR relaxation rates $1/T_1$ and $1/T_2$ and for
the uniform susceptibility.
I also discuss the transition from the spin liquid to antiferromagnet and the
critical behavior of these properties at the transition.
I compare these theoretical results with the data on high $T_c$
superconductors.
Finally, I discuss the formation of the spin gap due to the spin exhange
interaction between adjacent layers in bilayered materials.

\end{abstract}

In this talk I discuss the properties of an intermediate magnetic phase which
may be sandwiched between the Fermi liquid and antiferromagnet in strongly
correlated two dimensional electron systems.
If the concentration of electrons is low and the interaction between them is
weak, the electrons form a Fermi liquid.
If the band is half-filled and the interaction between electrons is strong the
system is an insulating antiferromagnet.
As the interaction between electrons and their density is varied the system
evolves from a Fermi liquid to an antiferromagnet.
There are at least two scenarios for this evolution \cite{disclaimer}:

\begin{itemize}
\item{There is just one transition leading directly from the Fermi liquid
to the antiferromagnet.
In the Fermi liquid near the transition line the antiferromagnetic spin
fluctuations have a small but non-zero gap.
At the energy scales larger than the gap the critical spin fluctuations lead
to non-Fermi liquid exponents in the Green function and spin response of the
electrons.
The low energy properties (at energies smaller than the antiferromagnetic
spin fluctuation gap)
of  the fermion quasiparticles are not qualitatively different from those
of the free electrons.
}

\item{The evolution from the Fermi liquid to the antiferromagnet goes through
the intermediate state as shown in Fig. 1.
In this intermediate state  Fermi liquid theory is not a correct
description of the low energy properties even far from criticality.
One theoretical realization of such an intermediate state is known as a ``spin
liquid'' \cite{spin-liquid}.
A spin liquid possesses a Fermi surface, spin $1/2$ fermionic excitations with
constant density of states at low energies and a particle-hole continuum but
the
fermions interact via a singular interaction mediated by the gauge field.
As discussed below the singular interaction causes anomalous temperature
dependence of the susceptibilities and NMR relaxation rates for a range of
values of density or interaction.
}

\end{itemize}

The evolution from the Fermi liquid to the antiferromagnet is realized
experimentally when the doping is varied in high $T_c$ materials.
As discussed in the preceding talk by A. Millis, the available
experimental data on the magnetic properties of these materials in the
``underdoped'' regime are incompatible with the theoretical predictions for
the critical behavior in the Fermi liquid scenario.
In this talk I will focus on the theoretical predictions \cite{AILM} for the
``spin-liquid'' scenario.
I shall outline the basic ideas of the theoretical reasoning leading to these
predictions first \cite{AIM}.

\begin{figure}
\vspace{-2.cm}
\centerline{\epsfxsize=6cm \epsfbox{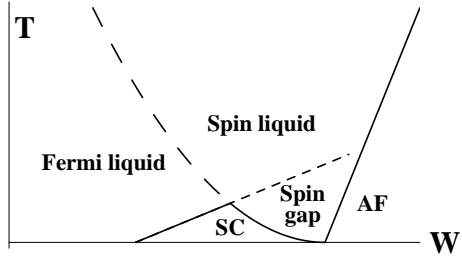}}
\vspace{-2.cm}
\nopagebreak
\caption{Phase diagram of a strongly correlated electron system in which
transition to the antiferromagnet happens through the intermediate spin liquid
phase.}
\label{f1}
\end{figure}

Our starting point for the description of the spin liquid is a Hamiltonian,
$H=H_{FL}+H_{gauge}$, where $H_{FL}$ describes fermions moving in a lattice
and interacting with each other via a  short range four fermion interaction
$W$:
\begin{equation}
H_{FL} = \sum_{p \alpha} \epsilon(p) c^\dagger_{p,\alpha} c_{p,\alpha} +
        W \sum_{p,p',q,\alpha,\beta} c^\dagger_{p,\alpha} c_{p+q,\alpha}
\label{H_FL}
\end{equation}
and $H_{gauge}$ describes the gauge field and its coupling to the fermions:
\begin{equation}
H_{gauge} = \sum_{p,k, \alpha} {\bf a}_k {\bf v}(p)
c^\dagger_{p-k/2,\alpha} c_{p+k/2,\alpha} +
	\frac{1}{2g_0^2} \sum_k ({\bf k} \times {\bf a})^2
\end{equation}
Here $g_0$ is the bare fermion-gauge field interaction constant, $\alpha=1,
\ldots N$ is a spin index.
In the physical spin liquid $N=2$, however, to obtain analytical results it is
convenient to consider limits $N \rightarrow 0$ and $N \rightarrow \infty$ and
interpolate between them.\cite{AIM}

Perturbation theory in $g$ leads immediately to two effects.
(i) The gauge field propagator becomes
\begin{equation}
D(\omega,k) = \frac{1}{\frac{Np_0|\omega|}{2\pi |k|} +\frac{1}{N^{1/2}g^2} k^2}
\label{D}
\end{equation}
Here the first term in the denominator is due to Landau damping of the gauge
field, $p_0$ is the curvature of the Fermi surface at the point where the
normal to the Fermi line is perpendicular to $\bf{k}$ and
$g^2 = N^{-1/2} g_0^2$.
(ii) Using the gauge-field propagator to calculate the self energy one finds
\cite{Lee}
\begin{equation}
\Sigma^{(1)}(\epsilon)=-i \left| \frac{\omega_0}{\epsilon} \right|^{1/3}
        \epsilon
\label{Sigma^(1)}
\end{equation}
where the energy scale $\omega_0$ can be expressed through  $g$, $p_0$ and
$v_F$.

{}From the gauge field propagator (\ref{D}) it is clear that typical scattering
involves typical mometum transfer $k \propto N^{1/2} \omega^{2/3}$.
Such processes change only slightly the direction ${\bf v}_F$ in which the
fermion propagates.
In the cartesian coordinates associated with ${\bf v}_F$ the Green function of
the fermion becomes
\begin{equation}
G^{(1)}(\epsilon,p) = \frac{1}{\Sigma^{(1)}(\epsilon) -
v_F [p_\parallel + p_\perp^2/(2p_0)]}
\label{G}
\end{equation}

Due to the large momentum transfer in the limit $N \rightarrow \infty$ the
curvature of the Fermi line becomes important and higher order terms of the
perturbation theory are small in $1/N$ \cite{AIM}.
This is reminiscent of the Migdal theorem in conventional electron-phonon
problem.
In alternative limit of $N \rightarrow 0$ the curvature of the Fermi line is
unimportant, the dependence of the fermion Green function on the momentum
$p_\perp$ can be neglected.
In this limit higher order diagrams become the same as in a one-dimensional
theory which can be solved by bosonization \cite{ILA}.
The results obtained by this method are qualitatively different from the
results of the $1/N$ expansion: the power laws predicted in the $1/N$
expansion (see below) are converted into more rapid exponential dependences.
However, the bosonization is only valid in the strict $N \rightarrow 0$
limit, at any fixed non-zero $N \ll 1$ the curvature terms become eventually
important.
We have obtained the leading behavior of the physical quantities
at fixed $N \ll 1$ using Ward identities.
It turns out that this behavior is qualitatively more similar to the behavior
at $N \gg 1$ than to exponentially rapid dependencies of $N \rightarrow 0$
limit.
Since the properties in the limits $N \gg 1$ and $N \ll 1$ are similar, but
the latter case is more technically difficult  I will mainly focus on the $N
\gg 1$ limit in this talk.

In this limit direct calculations show that at general wavelength $|{\bf q}|
\neq 2p_F$ response functions of the spin liquid are very
similar to  properties of the Fermi liquid \cite{AIM}
because of the small phase volume available for virtual processes which leave
both fermions with momentum  transfer $\bf p+q+k$ and  $\bf p+k$ close to the
Fermi surface.
The effects of the gauge field on the fermion vertices with large
momentum transfer are more interesting.
The situation changes drastically for $|q|$ close $2p_F$.
In this case a virtual process with momentum transfer $\bf q$ along the
Fermi surface leaves both fermions with
momenta $\bf p+q+k$ and $\bf p+k$ near the Fermi surface.
The leading  contribution in $1/N$ to the fermion spin fluctuation
vertex $\Gamma_{Q}$ is logarithmically divergent at $Q=2p_F$;
we find that higher powers of $N$ contain higher powers of logarithms;
we were able to sum these logarithms using a renormalization group method
obtaining \cite{AIM} power law singularities in $\Gamma_{2p_F}$:
\begin{equation}
\Gamma_{2p_F} \simeq \frac{1}{
        \left(\frac{\omega}{\omega_0}\right)^\sigma +
                \left(\frac{v_F k_{\parallel}}{\omega_0}\right)^{3\sigma/2} }
\label{Gamma_2pF}
\end{equation}
The exponent $\sigma$ can be calculated in the limits $N \rightarrow
\infty$ and $N \rightarrow 0$.
Extrapolation of the results obtained in these limits to the physical case
$N=2$ gives the estimates $1/4 < \sigma < 3/4$.

The singularity of the vertex means that the calculation of the partcle-hole
susceptibility must be reconsidered.
The change in the fermion Green function and the
singularity of the $2p_F$ vertex have profound effects on the fermion
polarization operator $\Pi(\omega,q)$.
This effect is especially interesting if $\sigma > 1/3$.
In this case, the polarization operator becomes singular at
$|{\bf q}|=2p_F$ ring in the momentum space \cite{AIM}:
\begin{equation}
\Pi ( \omega, {\bf q}) = \sqrt{\frac{\omega_0 p_0}{v_F^3}}
        \frac{1}{
        \left [ c_{\omega}
        \left ( \frac{|\omega|}{\omega_0} \right )^{2\sigma - 2/3}
        +  c_k \left(\frac{||q|-2p_F| v_F}{\omega_0} \right)^{3\sigma-1}
        \right ] }
\label{Pi2}
\end{equation}
The full susceptibility $\chi$ is obtained by combining the irreducible bubble
$\Pi$ with the short-range four fermion vertex $W$.
We have shown \cite{AIM} that the gauge-field interaction renormalizes a
sufficiently weak initial $W$ to zero, so $\chi(\omega,k) = \Pi(\omega,k)$
Thus, in this case, the susceptibility is singular in momentum space for a
wide range of $W$ and density.

The divergence of the susceptibility at $\omega=0$, $Q=2p_F$ results in a
strong temperature dependence of the NMR relaxation rates which are given by
summing the appropriate combinations of susceptibilities over momenta
\cite{AILM}:
\begin{equation}
\frac{1}{T_1T} \sim A^2 \frac{p_F p_0^{1/2} \omega_0^{1/2}}{v_F^{5/2}}
        \left( \frac{T}{\omega_0} \right)^{\frac{1}{3} - 2 \sigma}
\label{T1T}
\end{equation}
If $\sigma < 1/2$, the rate $T_2^{-1}$ is non divergent and if $\sigma > 1/2$,
\begin{equation}
\frac{1}{T_2}  \sim A^2 \frac{\sqrt{p_F p_0} \omega_0}{v_F^2 a}
        \left( \frac{T}{\omega_0} \right)^{1-2 \sigma}
\label{T_2}
\end{equation}
Here $a$ is the lattice constant.
For the uniform susceptibility we found
\begin{equation}
\chi_U = const + D_0^{\prime \prime} (T/ \omega_0)^{1+\beta}
\end{equation}
where $D_0''$ is a constant of order $\left( \frac{p_F g_e}{v_F} \right)^2 W$,
whose sign is positive for repulsive $W$ and negative for attractive $W$.
The exponent  $\beta (N) = \frac{4}{3} - \frac{1}{N}$.

NMR experiments on $La_{2-x}Sr_xCuO_4$ have shown that the copper $T_1$
rates has the temperature dependence
\begin{equation}
^{Cu}(T_1T)^{-1} \sim T^{-1}
\end{equation}
for $100K < T < 500K$ \cite{Kitaoka}.
The uniform susceptibility is given by $\chi \sim const + AT$ at least for
$150K <T < 400K$ \cite{MillisMonien}.
The strongly coupled spin liquid results with $\sigma \approx 2/3$ are in
agreement with these data.
However, the recent data show \cite{Walstedt} that $1/T_2$ in this material
scales as $1/(T_1T)$ in the temperature range $100-300 \; K$, which is not
consistent with the weaker dependence (\ref{T_2}).

Note that there is a fundamental difficulty in comparing quantitatively
the theoretical predictions with data.
All such predictions for the spin liquid or alternative
scenarios imply that an asymtotic low temperature behavior is reached.
At high temperatures non-universal properties such as details of the band
structure becomes important.
In fact, the existing data cover only a limited range of temperatures and can
be fit in different ways.
Therefore, it is very important that the measurements be extended to as low
temperatures as possible, especially on materials which do not display
spin gaps or superconductivity down to low temperatures.

Underdoped bilayered cuprates are known to exhibit a ``spin gap'' phenomena in
a broad temperature range $T_c<T<T_{sg}$ with $T_{sg} \approx
150 \; K$.
The theory of this phenomena in the spin liquid scenario is given by the
Cooper pairing of spinons on different planes \cite{AI}.
This pairing results from the antiferromagnetic interaction between the planes
$H= J_\perp \sum_i S^{(1)}_i S^{(2)}_i$ enhanced by long ranged spin
correlations in each plane \cite{JETP,Ubbens}.
Formally, in the regime of strong interaction with the gauge field the
effective interaction between spinons becomes dressed by the large vertices
(\ref{Gamma_2pF}) \cite{spin_gap}.
Due to the singular momentum dependence of the vertex  (\ref{Gamma_2pF}) the
gap equation is almost one dimensional, i.e. Cooper pairing occurs almost
independently on different parts of the Fermi line.
Therefore, the gap opens first only at a small patch of the Fermi line and
spreads to the rest of the Fermi line only at lower temperatures.
The energy of the gap is only weakly sensitive to the symmetry of the gap
function.
In the absence of local repulsion between spinons the $s$-state has somewhat
lower energy, but such repulsion, induced, e.g. by incoherent tunneling of
spinons from one plane to another, would lead to a lower energy of the
$d$-state symmetry.

Another direct application of the gauge theory discussed in this talk might be
$\nu=1/2$ Quantum Hall state in the $2D$ electron gas with screened Coulomb
interaction where charge response at $2p_F$ would be very singular.

\end{document}